# Goos–Hänchen-Shift Photonic Sensor for Nanometer-Scale Delayering and Tamper Detection in Semiconductor Packages

**MIA MOHAMMAD SHOAIB HASAN,[1] & MOHAMED ELKABBASH[2,3*]**

[1]*Bangladesh Army International University of Science and Technology, Comilla 3501, Bangladesh*
[2]*James C. Wyant College of Optical Sciences, University of Arizona, Tucson, Arizona, USA*
[3]*Department of Physics, University of Arizona, Tucson, Arizona, USA*

**melkabbash@arizona.edu*

**We propose a co-packaged photonic tamper sensor that detects progressive delayering and localized drilling through changes in the Goos–Hänchen (GH) shift of a reflected optical beam. Frustrated total internal reflection (FTIR) couples the beam into a high-index sensing layer, where its transverse-wavevector components acquire a thickness-dependent propagation phase. Numerical simulations show an approximately linear response for sensing-layer thicknesses below 250 nm and a delayering sensitivity of 5.3 nm of beam displacement per nanometer of material removed, nearly three times that of a conventional total internal reflection (TIR) structure. The off-resonant response remains stable under representative refractive-index, wavelength, and incidence-angle variations. Localized drilling also produces a monotonic GH-shift change that increases with drill depth and width. These results establish GH-shift readout as a rapid, spatially encoded, and difficult-to-emulate approach for semiconductor-package tamper detection.**

**Introduction.** Physical tampering of semiconductor packages and chips threatens the confidentiality and integrity of semiconductor systems by enabling attackers to expose internal circuitry, probe sensitive signals, extract cryptographic assets, or modify device operation [1]. Effective anti-tamper protection must therefore increase the attacker's *time-to-compromise* the chip and to shorten the *time-to-detection*, in order to trigger zeroization or shutdown of the sensitive devices[1-3]. In addition, the anti-tamper sensor should remain reliable under normal environmental and electromagnetic variation. Finally, and arguably most importantly, a sophisticated attacker should find it difficult to emulate or reproduce the measured signal because this increases the time-to-compromise and means that the sensing method is robust.

The current state of the art is the electronic active mesh where dense metal traces cover protected circuitry and carry randomized challenge patterns whose interruption, shorting, or rerouting triggers a tamper response[1, 4]. This approach provides rapid detection and increases attack difficulty because the adversary must preserve unpredictable electrical continuity while accessing the circuit. Commercial implementations are available, but they remain vulnerable to sophisticated FIB-assisted removal, reconstruction, or bypass strategies[5]. Furthermore, electronic meshes only protect the physical region they occupy and only provide die-level anti-tamper security.

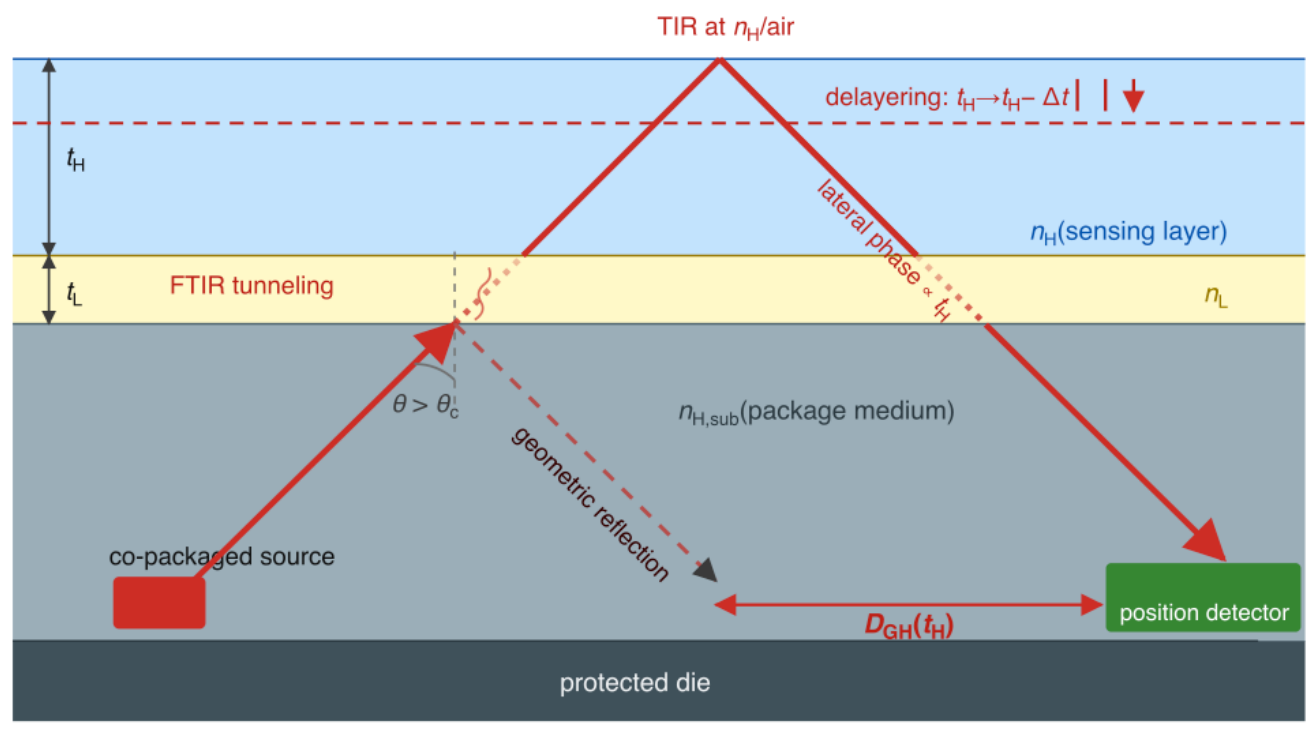

**Fig. 1.** Schematic of the co-packaged GH-shift delayering

Moreover, advanced packaging substantially expands the physical attack surface by integrating multiple chiplets, interposers, redistribution layers, through-silicon vias, and die-to-die interconnects within a single package[6]. An attacker may therefore enter through the package, target an individual chiplet, or access the interposer and communication links, creating a need for continuous security from the package boundary to the protected dies and across multiple components[7]. Recent package-integrity approaches have used indirect thermal, electrical, and environmental signatures. These mechanisms add little physical delay, and can sometimes be spoofed by reproducing expected environmental conditions [8].

Because any front-side invasive access attack would require delayering the encapsulation to expose the die or interconnects for imaging, microprobing, or FIB modification, and to reduce the material barrier for optical or laser-based attacks, a sensor that detects nanometer scale delayering can provide package-to-die level anti-tamper protection. To our knowledge, no current protection method provides rapid, nanometer-scale sensing of thin-film delayering while generating an optical signature that is difficult for an attacker to emulate.

Here, we introduce a Goos–Hänchen (GH)-shift-based delayering and thin-film drilling sensor for package and chip anti-tamper protection. The GH shift is the lateral displacement of a finite optical beam upon TIR [9, 10]. It arises because the beam contains a spectrum of in-plane wavevector components $k_x$ , each of which acquires a different reflection phase; their recombination shifts the reflected beam laterally [9, 10]. The GH shift has been investigated as a refractometric sensing mechanism because changes in the surrounding refractive index modify the reflection phase and therefore the lateral position of the reflected beam [11-13]. In this work, we introduce a co-packaged GH-shift photonic sensor for detecting both progressive delayering and localized drilling of semiconductor packages. While GH shift in resonant systems is highly sensitive, this sensitivity trades off with robustness to environmental and mechanical instabilities that can lead to a false positive signal which triggers the chip destruction. We show that we can increase the sensitivity without trading-off robustness to environmental and mechanical perturbations through using FTIR-based GH shift. FTIR couples the incident beam into a high-index sensing layer, producing a linearly dependent GH shift on the high-index layer's thickness as long as the thickness is below the Fabry-Perot condition (here ≈ 250 nm). This approach enhances the delayering sensitivity by nearly threefold relative to a conventional low-index TIR structure while remaining robust against environmental and source variations, including thermo-optic refractive-index changes, wavelength drift, and incidence-angle deviations.

**Proposed sensor.** The proposed sensor is schematically shown in Fig. 1. The proposed co-packaged optical security architecture has an integrated laser or VCSEL that launches a finite Gaussian beam into a high-index package layer at an oblique angle. An on-package detector array records the reflected beam. Because a finite beam contains a distribution of in-plane wavevector components $k_x$, each component experiences a complex reflection coefficient $r(k_x) = |r(k_x)| e^{i\phi(k_x)}$. The $k_x$-dependent reflection phase causes the reflected components to recombine with a lateral displacement, giving the Goos–Hänchen shift $D_{\mathrm{GH}} \approx -\partial\phi/\partial k_x$. Removal of the package material changes the multilayer complex reflection coefficient and therefore modifies the GH shift which affects the beam centroid and the transverse intensity distribution. After calibration, these optical observables can be directly measured through a CMOS photodetector array to trigger shutdown or zeroization before the protected circuitry is exposed.

The proposed GH sensor offers strong security against delayering attacks. To emulate the untampered response after delayering, an attacker would need to remove or redirect part of the altered reflected beam and inject an additional beam with the same spatial intensity distribution required so that the combined profile matches the calibrated untampered profile. Where the tampered intensity exceeds the target profile, light must first be attenuated or leaked out. On the other hand, where the tampered intensity falls below the target, correctly shaped light must be added. Reproducing the complete spatial profile is therefore substantially more difficult than spoofing a single scalar sensor output. The difficulty in spoofing the signal lengthens the time-to-compromise based on a delayering attack. Moreover, the time-to-detect from a position sensitive photo-detector can be on the order of 10 ns, although the total detection latency will depend on the readout electronics [14].

We first consider a simple GH-shift sensing scheme where TIR occurs at the interface between a high-index medium enclosing the source and a low-index thin-film (**FIG. 2a** inset). **FIG. 2a** shows a finite element method analysis (COMSOL Multiphysics) on this structure. A Gaussian beam with an incident angle of 65° was launched from the bottom boundary of the prism toward the low index layer. The thickness of low index layer was varied from 200 nm to 20 nm to simulate a physical packaging delayering scenario. Under these conditions, the beam generates an evanescent field that penetrates into the low-index film and decays before reaching the outer medium, so the structure behaves close to a single TIR boundary with a weak perturbation from the finite film thickness. As the thickness of the low-RI layer increases from 20 nm to 200 nm, the GH shift changes between ~ 0.76 μm to ~1.1 μm, with a total shift of 325 nm leading to a delayering sensitivity of 1.8 nm beam shift per 1 nm material removal.

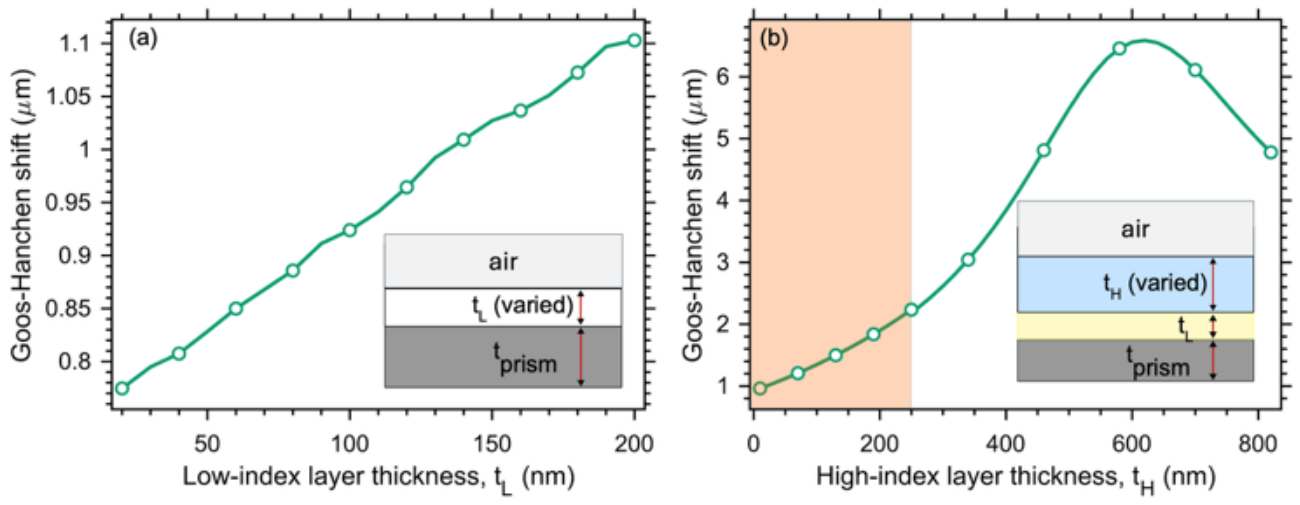

**Fig. 2.** GH shift as a function of **(a)** low-index-layer thickness and **(b)** high-index sensing-layer thickness.

Adding a high-index layer above the low-index film enables FTIR, allowing the evanescent field to couple into a propagating mode in the high-index layer before undergoing TIR at the high-index/air interface. Each transverse-wavevector component $k_x$ then accumulates a round-trip propagation phase $\phi_{\mathrm{prop}} = 2k_{z,H}t_H$, where $k_{z,H} = \sqrt{n_H^2 k_0^2 - k_x^2}$. Because $k_{z,H}$ depends on $k_x$, this propagation adds a thickness-dependent contribution to the reflection-phase gradient and therefore to the GH shift. In the off-resonant regime, the corresponding displacement is $D_{\mathrm{prop}} = 2t_H k_x / k_{z,H} = 2t_H \tan\theta_H$, giving an approximately linear dependence on $t_H$. Consequently, the GH shift increases from 0.96 to 2.24 μm over the investigated linear range (~ 250 nm) [**Fig. 2(b)**], corresponding to a delayering sensitivity of 5.3 nm of beam displacement per nanometer of material removed, nearly three times that of the low-index TIR structure.

The minimum detectable delayering depth is determined by the smallest resolvable change in the reflected-beam centroid divided by the delayering sensitivity. Given a beam-position accuracy of approximately 150 nm as a benchmark[15], the low-index TIR sensor structure, with a sensitivity of 1.8 nm of beam displacement per nanometer of material removed, yields an estimated detection limit of ≈ 83 nm. In comparison, the FTIR-enhanced high-index structure, with a sensitivity of 5.3 nm/nm, yields a detection limit of ≈ 28 nm.

The linear regime is selected as the primary operating range for tamper detection because it provides a simple, stable, and directly invertible relationship between the measured beam displacement and the remaining layer thickness. In this regime, $D_{\mathrm{GH}} = D_0 + S_t t_H$, so each measured displacement corresponds uniquely to $t_H = (D_{\mathrm{GH}} - D_0)/S_t$, allowing the amount of removed material to be determined without nonlinear calibration. **Figure 2b** shows the GH shift over the full $0 - 820$ nm high-index-layer thickness range, with an approximately linear response below $t_H \approx 250$ nm. The response remains monotonic and therefore single-valued beyond this region, up to the Fabry–Pérot resonance peak near $t_H \approx 600$ nm; thus, this pre-resonant nonlinear range could also be used after calibration. However, operation closer to resonance is more sensitive to refractive-index, wavelength, incidence-angle, and temperature variations, as we show in the next section.

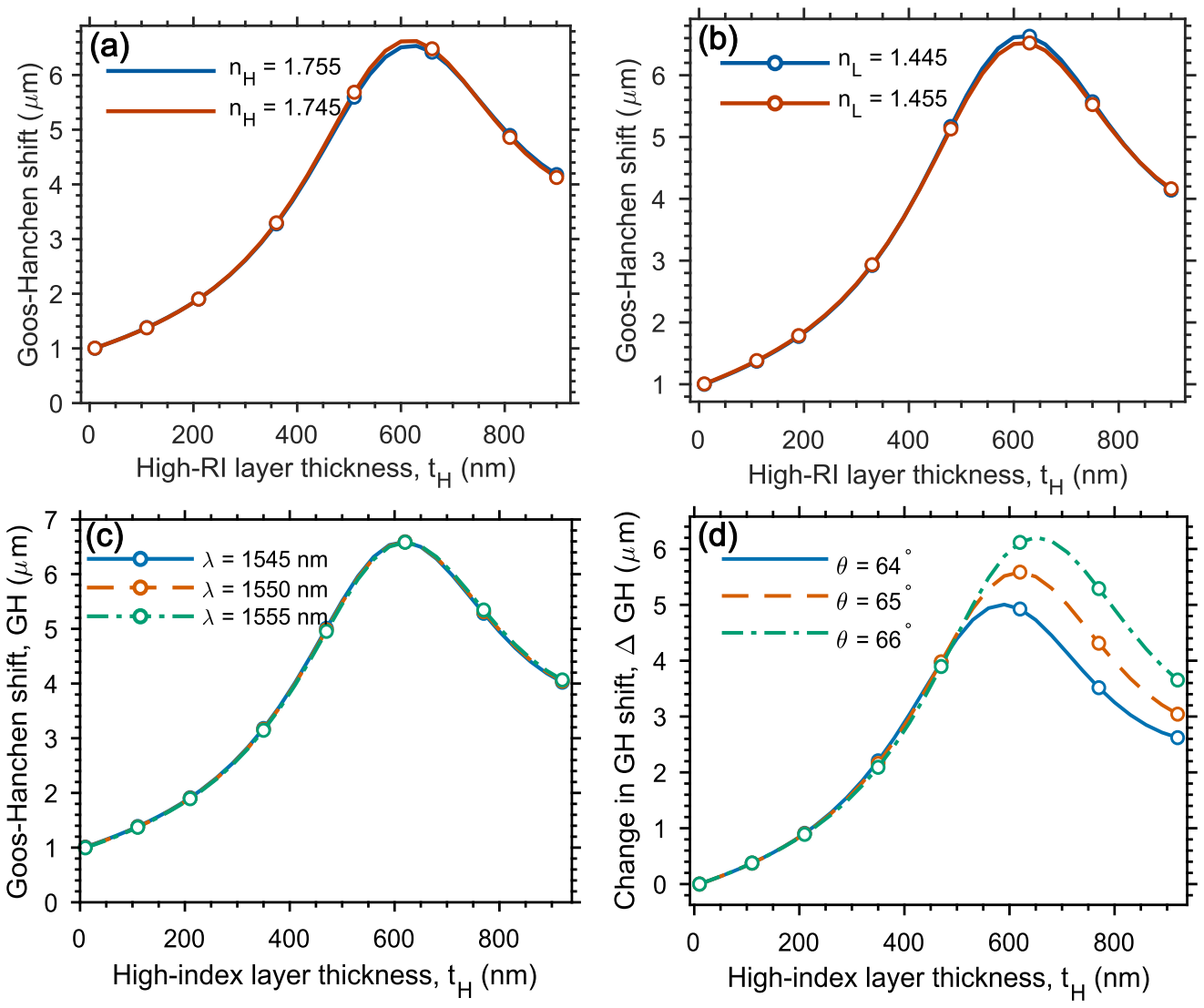

**Fig. 3.** Goos–Hänchen shift as a function of high-index-layer thickness for variations in **(a)** high-index refractive index, **(b)** low-index refractive index, **(c)** source wavelength, and **(d)** incidence angle.

**Robustness to Environmental and Optical-Source Variations.** Environmental robustness is a critical requirement for package anti-tamper sensors because operating variations can alter the measured signature and trigger false alarms, alternatively if the detection threshold is widened, this would reduce sensitivity to actual attacks [8]. This problem is particularly important for methods that infer package integrity from environmentally sensitive quantities such as temperature, thermal dissipation [8]. The anti-tamper literature therefore identifies a low false-positive rate under legitimate conditions as a central design objective and explicitly evaluates temperature, humidity, vibration, and internal component motion as potential sources of measurement drift. Motivated by these requirements, we evaluate the environmental robustness of the proposed GH-shift sensor against variations in the refractive indices of both dielectric layers, incidence angle, and source wavelength.

**Refractive index tolerance: Fig. 3(a)** and **Fig. 3(b)** evaluate a total refractive-index variation of $\Delta n = 0.01$ for both dielectric layers. The representative material for the high-index layer, *is* $Al_2O_3$ with a thermo-optic coefficient of ~ $dn/dT = 2.75 \times 10^{-5}\ \mathrm{K}^{-1}$ [16]. Accordingly, the simulated range corresponds to a total temperature excursion of approximately ≈ 364°C. $\mathrm{SiO_2}$ is representative of the low-index layer with a thermo-optic coefficient ~ $1.1 \times 10^{-5}\ \mathrm{K}^{-1}$ [17]. The simulated range corresponds to a total temperature change of approximately 900°C. The change in the GH-shift response is negligible in both cases.

**Wavelength instability tolerance:** A 1550-nm co-packaged VCSEL may undergo a temperature-induced wavelength shift of several nanometers during normal operation; using a representative coefficient of 0.11 nm/K, a $25 - 85$°C temperature change corresponds to ~ 6.6 nm[18]. We therefore evaluated a conservative wavelength range of $1540 - 1560$ nm while holding the incidence angle fixed. As shown in **Fig. 3(c),** the GH-shift curves nearly overlap throughout the linear operating regime, indicating negligible wavelength-induced error in the thickness calibration. The dominant thickness-dependent contribution is the geometric propagation displacement, $D_{\mathrm{prop}} = 2t_H \tan\theta_H$, which is wavelength-independent for fixed angle and refractive indices. Residual wavelength dependence arises from the TIR and FTIR interface phases and from the wavelength-dependent optical thickness of the multilayer. These effects remain weak over the investigated range, with only small differences appearing near the Fabry–Pérot resonance, where the phase response is more dispersive.

**Beam alignment tolerance:** A single co-packaged source can be combined with a diffractive optical element to generate multiple interrogation beams that illuminate different regions of the package[19], each at an angle satisfying the TIR condition and each mapped to a corresponding photodetector. This architecture increases spatial coverage while retaining a compact source footprint and enables simultaneous monitoring of multiple access regions. During operation, however, the effective incidence angle may vary due to source or grating thermo-mechanical deformation, or wavelength-dependent diffraction. For a grating designed to redirect the beam at 65°, a wavelength drift of 6.6 nm corresponds to an angular change of approximately 0.52°. The sensing response must therefore remain stable over a realistic angular range so that benign angular drift does not produce a false delayering signal. Incidence-angle tolerance was evaluated by varying the source angle from 64° to 66° around the nominal value of 65°. As shown in **Fig. 3(c),** the GH-shift curves nearly overlap throughout the linear operating range, indicating that a ±1° alignment variation *produces no significant change* in the thickness-to-displacement calibration. Small angle-dependent differences become apparent only at larger high-index-layer thicknesses, where the Fabry–Pérot resonance is approached. In this regime, changing the incidence angle modifies the in-plane wavevector and therefore shifts the resonance condition, changing both the peak GH shift and its corresponding thickness. The proposed off-resonant linear operating regime is therefore substantially more tolerant to source-angle variations than the resonant regime.

**Detection of Localized Drilling Attacks.** Our analysis, so far, has considered spatially uniform delayering of the sensing layer. In practice, however, an attacker may instead use localized drilling or milling to create a narrow access path through the package while leaving most of the surface intact. To determine whether the proposed sensor can detect this more localized form of material removal, we calculated the relative GH-shift change as a function of drill depth for several drill widths. The response increases monotonically with drill depth for all widths, while wider openings produce substantially larger shifts because they perturb a greater fraction of the reflected beam. Consequently, wider drills reach a given detection threshold at smaller depths, whereas narrower drills require deeper penetration before becoming detectable. These results show that the sensor is sensitive not only to uniform delayering but also to localized material-removal attacks.

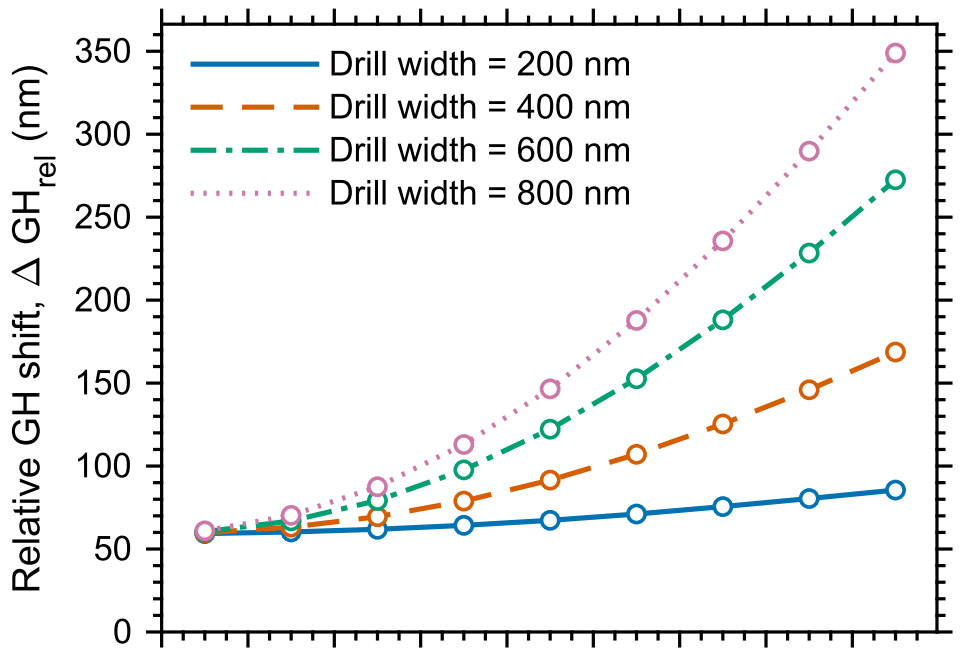

**Fig. 4**. Relative GH-shift change versus drill depth for localized openings of different widths, showing stronger and earlier detection for wider drill features.

**Conclusion and outlook.** We introduced the GH shift as a mechanism for environmentally robust detection of semiconductor-package tampering. The tolerance analysis mitigates key concerns associated with co-packaged optical implementation. Unlike electronic or optical-waveguide meshes, which protect only the regions traversed by the mesh, the proposed approach can interrogate extended package areas. Combining this distributed package-level sensor with localized die-level optical sensors, such as BEOL plasmonic resonators sensitive to refractive-index modification, could provide simultaneous package-to-die protection and extend detection to FIB-scale attacks with feature sizes of a few tens of nanometers[20].

**Disclosures**. The authors declare no conflicts of interest.

**Data Availability.** Data are available upon request from the authors.